\documentclass[3p,times]{elsarticle}
\usepackage{amssymb}
\usepackage[figuresright]{rotating}
\usepackage{epsfig}
\usepackage{epstopdf}
\usepackage[T1]{fontenc}
\usepackage{amssymb}
\usepackage{amsmath}
\usepackage{amsfonts}
\usepackage{verbatim}
\usepackage{graphicx}
\usepackage{hyperref}
\usepackage[usenames,dvipsnames]{xcolor}
\usepackage{algorithm}
\usepackage{array}
\usepackage{subcaption}
\usepackage{caption}
\usepackage{pgfplots}
\usepackage{natbib}
\usepackage{enumerate}
\usepackage{colortbl}
\usepackage{enumitem}

\usepackage{algpseudocode}
\author{Paresh Baidya$^1$, Swagata Mondal$^2$, Rourab Paul$^3$, 
}
\address{Department of CSE, Siksha O Anusandhan (Deemed to be University), Bhubaneswar, Odisha India Siksha O Anusandhan, Bhubaneswar, India$^{1}$, \\ 
Dept. Electronics and Communication, Jalpaiguri Government Engineering College,  Maulana Abul Kalam Azad University $^2$ \\  
Dept. of Computer Science, University of Pisa, Italy$^3$\\ 
pareshbaidya@soa.ac.in$^1$, swaga89@gmail.com$^2$, rourab.paul@unipi.it$^3$ 
}
\begin{document}
\title{Near Threshold Computation of Partitioned Ring Learning With Error (RLWE) Hardware Accelerator on Reconfigurable Architecture}
\begin{abstract}
The Ring Learning With Error (RLWE) algorithm plays a crucial role in Post Quantum Cryptography (PQC) and Homomorphic Encryption (HE). The security of existing classical crypto algorithms is reduced in quantum computers. The adversaries can store all encrypted data, and once quantum computers become available, they can potentially expose this encrypted data. To address this emerging threat, researchers and cryptographers are actively developing and exploring quantum-resistant cryptographic algorithms like RLWE. On the other hand, the HE allows operations on encrypted data which is appropriate for getting services from third parties without revealing confidential plain-texts. The Field Programmable gate Array (FPGA) based PQC and HE hardware accelerators like RLWE is much cost-effective than processor based platform and Application Specific Integrated Circuit (ASIC). FPGA based hardware accelerators still consume more power compare to ASIC based design. Near Threshold Computation (NTC) may be a convenient solution for FPGA based RLWE implementation. This paper implements RLWE hardware accelerator which has 14 subcomponents. This paper creates clusters based on the critical path of all 14 subcomponents. Each cluster is implemented in an FPGA partition which has the same biasing voltage $V_{ccint}$. The clusters that have higher critical paths use higher $V_{ccint}$ to avoid timing failure. The clusters have lower critical paths use lower biasing voltage $V_{ccint}$. Any timing error causes fr NTC can be caught by the RAZOR flipflop used in each subcomponents of RLWE. This voltage scaled, partitioned RLWE can save $\sim$6\% and $\sim$11\% power in Vivado and VTR platform respectively. The resource usage and throughput of the implemented RLWE hardware accelerator is comparatively better than existing literature.
\end{abstract}
\maketitle
\section{Introduction}
\label{sec:intro}
Lattice based cryptography is currently considered as one of the most secure solutions compared to classical cryptography schemes. The discrete logarithm (Elliptic Curve Cryptography), RSA and ECDSA classical schemes are used to secure modern Internet communications. These asymmetric key crypto-systems are based on the hardness of prime factor and discrete logarithm. However, asymmetric key crypto-systems are no longer secure under quantum attacks. Even in brute force, Grover's quantum algorithm \cite{grover} reduces the searching space of symmetric key cryptography: Advanced Encryption Scheme (AES) from $O(2^{n})$ to $O(2^{n/2})$, where $n$ is the key size. Therefore the security AES-256 is comparable to AES-128 in quantum computer. However, no such computationally strong quantum computer has been developed yet. IBM and Google have claimed to develop such computationally extensive quantum computers within a few years. Post-quantum crypto algorithms generally deal with non-quantum operations but it strongly resists both classical and quantum attacks. Mainly there are four post quantum cryptography schemes are avialable: (i) Code based cryptography, (ii)Lattice based cryptography, (iii)Hash based cryptography and (iv)Multivariate quadratic cryptography. Among all these schemes lattice based cryptography(LBC) is computationally efficient. 
On 2005, O. Regev \cite{regev11} first introduced a lattice based cryptography (LBC) scheme name as Learning With Error (LWE). Later LBC becomes more popular for its significant theoretical progress \cite{regev11,micciancio2009}. The LBC becomes more suitable for real life applications when it was implemented in software and hardware platform by articles \cite{pop,gottert2012design,poppelmann2014,sinhaKnuth,aysu2013low}. On the other hand, post-quantum cryptography finds its application in more advanced schemes such as fully homomorphic encryption, which allows the operations on the encrypted data without revealing any information to the third party.
\par
For hardware implementation of RLWE as PQC and HE, designer can choose two platforms: FPGA or ASIC.
In practical, FPGA based PQC and HE hardware accelerators like RLWE is much cost-effective compared to ASIC. However, FPGA based hardware accelerator is less energy efficient compared to ASIC \citep{hybrid} because of its programmable switch. An experimental study  \citep{Kusse} shows power consumption of 8 bit full adder in  Xilinx XC4003A  is 100 times more compared to CMOS ASIC platform. The implementation of 8-bit adder in FPGA platform consumed the power 4.2mW/MHz at 5V and the same implementation in ASIC consumed 5.5uW/MHz at 3.3V \citep{Kusse}. The power consumption of FPGA based design is crucial specially for battery powered embedded system. NTC may be an appropriate solution to reduce the power consumption of FPGA based RLWE implementation.
\par
To the best of our knowledge, all the existing RLWE implementations \cite{gottert2012design,Howe,pop,roy2014compact} focus on speed and resource optimization whereas, this is the first reported work which tries to reduce power consumption of RLWE.

\subsection{Related Work}
O. Regev \cite{regev11} introduced the first hardness proof of LWE cryptography in 2005. Later, various hardware and software implementation of LWE were proposed. There are many literature which implemented various subcomponents of RLWE algorithm. The polynomial multiplication, polynomial division and gaussian sampler used to generate error polynomials are the most challenging subcomponents of RLWE. Howe et. al \cite{Howe} proposed a hardware architecture for a standard lattice based cryptography (LWE) on a Spartan-6 FPGA platform in 2016. The primary contribution of this paper is area optimization of Gaussian Sampler by balancing  area and performance. Poppelman et al. \cite{poppelmann2014enhanced} reported a hardware implementations of Gaussian sampler of RLWE accelerator. They implemented Cumulative Distribution Table (CDT) based Gaussian sampler on reconfigurable hardware. Authors also compared CDT based Gaussian sampler and Bernoulli sampler. Roy et al. \citep{sinhaKnuth} implemented Gaussian sampler for sampling from the discrete gaussian distribution using Knuth Yao algorithm. Poppelman et al. \citep{poppelmann2012} implemented a polynomial multiplier which stores all twiddle factors in a dedicated memory required for NTT computation. Article \cite{gottert2012design} reduced the key size of the conventional LWE schemes and implemented a compact and efficient LWE hardware. The article \cite{gottert2012design} compared LWE hardware and software implementations with Lindner and Peikert's design \cite{lindner}. Article \cite{gottert2012design} introduced LWE-matrix and LWE-polynomial and compared these variants of LWE. Article \cite{gottert2012design} discarded the idea of LWE-matrix because of its large area consumption, they have only implemented RLWE-polynomial. Poppelman et al. \cite{pop} implements an efficient hardware of RLWE on Virtex 6 FPGA which can fit on a low-cost Spartan-6 FPGA. This hardware accelerator improved the work of \cite{gottert2012design}. Article \cite{poppelmann2014} proposed a lightest RLWE encryption scheme by reducing the resource, compared with the high speed implementation of \cite{pop}. 
Roy et al.\cite{roy2014compact} implemented a compact ring-LWE coprocessor on Virtex 6 FPGA where they optimized NTT polynomial multiplication and reduced the computation cost of the twiddle factors by avoiding the pre-computation overhead. A variation of the Native Title Research Unit (NTRU) Encrypt system with a quantum security reduction is used in article \cite{Damien}. Its security is based on the standard model's LWE problem in polynomial rings. All the above mentioned RLWE implementations have mostly focussed on area versus speed issue. More speed causes more area and more area causes more power consumption. Without affecting the datapaths (area vs speed), the conventional low power solutions mostly deal with clock gating, power gating, dynamic voltage and frequency islands, retention power gating,  save and restore power gating architectures. To the best our knowledge. this is the first RLWE hardware accelerator based on dynamic voltage island concept where different voltage islands run with near threshold voltages

\subsection{Proposed RLWE} The literature of RLWE competes the growing high speed applications by increasing throughput of the architecture. The throughput is mostly achieved by more parallelism which causes more power consumption. Power consumption is a severe concern for embedded systems but existing literature has not explored the possibilities of NTC in any PQC and HE implementations. To the best of our knowledge our proposed RLWE is the first NTC in FPGA and ASIC platform. In our implementation we designed a RLWE dedicated hardware which has 14 subcomponents. The critical path of all these subcomponents are measured and based on the critical path, 4 cluster algorithms create groups with these 14 subcomponents. The FPGA is partitioned and different partitions has different biasing voltage $V_{ccint}$. The group having higher value of critical path is placed in a partition which have higher $V_{ccint}$ and the group having lower value of critical path is placed in a partition which have lower $V_{ccint}$.   
The primary contribution of our paper is stated below:
\begin{itemize}
\item This paper designs and implements RLWE crypto algorithm in Artix-7 (xc7a100tcsg324-3,28nm) commercial FPGA and other 3 different academic FPGAs. The implmented RLWE hardware accelerator has 14 subcomponents such as Random Number Generator (RND), Poly\_div, NTT Controller, Polynomial Multiplier, Read only Memory (ROM), Scanner, Datapath Controller, Convolution Controller, Poly\_add, reader, Distance, Row Column Controller and 2 dual port Random Access Memory (RAM).
\item The cluster algorithms create groups of subcomponents which have similar critical paths. Different groups are placed in different FPGA partitions. Each partition runs with different $V_{ccint}$. The groups with lower critical path are placed in a lower $V_{ccint}$ partition and the groups with higher critical path are placed in a higher $V_{ccint}$ partition. The timing errors causes from the reducing $V_{ccint}$ are handled by the Razor Flipflops.
\item The proposed FPGA based voltage scaled RLWE can save ~11\% power consumption in VTR tool. The throughput, resource and power consumption of proposed RLWE is reasonably better compared to existing RLWE.   
\end{itemize}
The rest of the paper is organised as follows: Sec. \ref{sec:rlwe} states the background of the RLWE algorithm followed by Sec. \ref{sec:hw} which gives a preliminary idea about hardware implementation of RLWE. Next, in Sec. \ref{sec:tool}, the
The tool flow of the proposed model is described. In
Sec. \ref{sec:cluster}, the four cluster algorithm used to partition the FPGA is discussed and in Sec. \ref{sec:algo} and Sec. \ref{sec:rai}, the proposed algorithm to calculate biasing voltage and results
are stated respectively. Finally, Sec. \ref{sec:con}. concludes the paper

\section{The Ring-LWE Schemes}
\label{sec:rlwe}
 The Learning With Error (LWE) was proposed by O Regev in \cite{regev11} 2005. The LWE become popular because it is developed in secure lattice based cryptosystem which resists quantum attacks. Later LWE achieves computational efficiency and it reduces the key size by  adopting polynomial Ring \cite{lyubashevsky1}. The LWE schemes perform in $R_{q}= Z_{q}[x]/<f>$, where $f$ is irreducible polynomial of degree $n-1$, and $n=2^{k} \geq 1$ and a prime $q$ is chosen such a way that $q \equiv~1~mod~2n$. There is an error distribution called discrete Gaussian distribution $\chi_\sigma $ with the standard deviation $\sigma>1$ and mean 0. The LWE scheme defined as follows: 
\begin{itemize}
\item KeyGEN($a$): choose $r_{1}$,$r_{2} \in R_{q}$ sampled from $\chi_\sigma $ . Let $p=r_{1}-a\ast r_{2}$. The public key is $(a,p)$ and private key is $r_{2}$.The polynomial $a \in R_{q}$ chosen uniformly during the key generation.
\item EnCrypt($a,p,m$): choose error polynomials $e_{1},e_{2},e_{3} \in R$ sampled from $\chi_\sigma $. Then compute the cipher text as a polynomial
\begin{equation}
\label{equ:c1}
c_{1}=a\ast e_{1}+e_{2}
\end{equation}
\begin{equation}
\label{equ:c2}
c_{2}=p\ast e_{1}+e_{3}+ \tilde{m}
\end{equation}
where $\tilde{m} is encode(m)$ and $c_{1}$ and $c_{2}$ are the encrypted message.
\item DeCrypt($c_{1},c_{2},r_{2}$): Decryption process requires to compute below equation.
\begin{equation}
\label{equ:c3}
m'=c_{1}\ast r_{2}+ c_{2}
\end{equation}  
Thereafter, to get the original message $m$, it needs to decode $m'$.
\end{itemize}
\section{Hardware Architecture of RLWE}
\label{sec:hw}
The proposed $RLWE$ has 14 components such as Random Number Generator (RND), Polynomial Divider (Poly\_Div), NTT Controller, Polynomial Multiplier, Polynomial Subtraction (Sub\_mod), Scanner, Datapath Controller, Convolution Controller, Polynomial Adder (Poly\_add), reader, Distance, Row Column Controller and 2 dual port Random Access Memory (RAM).The main components in the architecture are: the memory file, Gaussian sampler, random number generator,NTT controller and datapath controller. The brief description of all hardware components are stated below:
\subsection{Random Number Generator (RND)}
This paper uses Trivium Random \cite{trivium} number generator to generate error polynomial $e_3$. In our application the secret key required for the RLWE is predefined. In real application the secret key should be generated randomly which can be done by the Trivium Random Number Generator. Trivium require one key and initialize vector to generate random numbers. 
\subsection{Polynomial Divider (poly\_div)}
The polynomial divider is required to divide polynomials. The resultant value of all the arithmetic operations stated in equ. \ref{equ:c1}, equ. \ref{equ:c2} and equ. \ref{equ:c3} need to bring under $R_{q}= Z_{q}[x]/<f>$.

\subsection{Dual Port RAM} This RAM stores the public key, message, gaussian sample value required for RLW execution. The dual port memory architecture allows parallel reading and writing operation of polynomial coeffcients. The RLWE needs to handle huge number of polynomial coefficients. The parallel read write operation of these polynomial coefficients increases the RLWE throughput significantly.
\subsection{Datapath Controller}
The Datapath Controller has three sub components such as:$Polynomial~Multiplier$, $Polynomial~Adder (poly\_add)$ and $Polynomial Subtraction (Sub\_mod)$ which are used to multiply, addition and subtraction of the polynomials.
\subsubsection{Polynomial Multiplier}
The polynomial multiplier (poly\_mul) is required to multiply $a$ and $e_{1}$ in equ. \ref{equ:c1} and $p$ and $e_{1}$ in in equ. \ref{equ:c2}. The implemented polynomial multiplier is based Fast Fourier Transform (FFT) which has the lowest timing complexity among all available algorithm for polynomial multiplier.
\subsubsection{Polynomial Adder (poly\_add)}
The polynomial Adder (Poly\_add) is required to add polynomials $a\ast e_{1}$ and $e_{2}$ in equ. \ref{equ:c1} and $p \ast e_{1}$, $e_{3}$ and $\tilde{m}$ in equ. \ref{equ:c2}. In this implementation we used usual process of polynomial addition. This block reads coefficients of polynomials from RAM and add them one by one.
\subsubsection{Polynomial Subtraction (Sub\_mod)}
The Polynomial Subtraction (Sub\_mod) is required to subtract polynomials. This block is required when Polynomial Divider (poly\_div) will perform. The  working principal of polynomial subtraction is similar to polynomial addition. The negative number is complemented here.
\subsection{Convolution Controller}
The convolution controller decides whether addition or multiplication will be executed.
This block used to control the Datapath Controller.
\subsection{Reader}
The reader block perform loading operation to load the coefficients of polynomials and Gaussian noise coefficients. It also encode the message bit before make into cipher text.
\subsection{NTT Controller} In RLWE scheme, polynomial multiplication is one of the the main operation of the encryption and decryption process. hardware implementation of coefficient wise multiplication of two polynomial functions is computationally expensive. Therefore, we convert the coefficient wise representation to point wise representation using the Number Theoretic Transformation (NTT). $NTT Controller$ performs the NTT operation on the polynomial functions  $a, p, r_{2}, e_{1}, e_{2}, e_{3}$ to generate $\tilde{a}$, $\tilde{p}$, $\tilde{r_{2}}$, $\tilde{e_{1}}$, $\tilde{e_{2}}$, $\tilde{e_{3}}$.
\subsection{Gaussian Sampler:scanner, distance, row column controller}
The Gaussian Sampler is based on Knuth Yao Sampler \cite{knuth}. The Knuth Yao Sampler is a tree based sampler which stores the binary expansion of the samples in a probability matrix. This probability matrix stores probabilities of the samples. Roy et al.\cite{sinhaKnuth} implemented the first hardware of Gaussian sampler using  Knuth Yao algorithm which used pre-computed table of the probability matrix. Knuth Yao algorithm uses a random walk model for sampling process. This sampling process creates a discrete distribution generating (DDG) tree using the probability matrix. Our $Gaussian~Sampler$ has three subcomponents such as: $scanner$, $distance$, $row~column~controller$ modules which are dedicated for traversing the tree, scanning the bit from the ROM and calculate the distance in the tree of visited node and intermediate node. The detail of the Knuth Yao sampler and DDG tree are describe in  \cite{knuth}.
\subsubsection{Scanner} Scanner block performs the scanning operation on the ROM block. It scans the bits from a ROM word. When all bits are read from a ROM word then it fetch next ROM word to scan.
\subsubsection{Row ColumnController} Row ColumnController has two registers: an up counter named as $column\_length$ and a down counter named as $row\_number$. The $column\_length$ stores the length of the different column length of the probability matrix. At the first step of column scanning process, $row\_number$ is initialized by the $column\_length$. If the $row\_number$ reaches to zero, the column scanning process is completed.
\newline
\subsubsection{Distance} The distance block requires to construct the DDG tree of the probability matrix. For this purpose a subtracter is used to control the random walk. When the subtracter value is $<0$, then it indicates completion of the sampling operation. After the completion of the sampling operation, the $row~column~controller$ selects current $row\_number$ as a sample output.  
 \section{Tool Flow}
 \label{sec:tool}
 This paper uses two environments 1)Python-Vivado and 2) Python-VTR.
 In Python-Vivado environment, the vivado generates the synthesis report which includes timing report of RLWE. The timing report consists of all timing related information of RLWE including the critical path length of all design paths. This timing report is sent to python environment to cluster the 14 subcomponents of RLWE using cluster algorithms mentioned in Sec. \ref{sec:cluster}.  This python environment generates the Xilinx Design Constraints (XDC) file which mentions coordinates of all 14 subcomponent of RLWE in FPGA floor. This coordinates of RLWE subcomponents are generated based on the partition or cluster. This XDC then includes in the implementation process of of $Vivado$ environment. The $Vivado$ used Artix 7, 28 nm FPGA. The VTR \cite{vtr} also uses the same process as mentioned for $Vivado$. VTR generates Synopsys Design Constraint (SDC) file instead of XDC. The VTR uses 3 academic FPGAs : 22nm, 45nm and 180nm.
\section{Cluster Algorithms}
\label{sec:cluster}
This paper analyses four cluster algorithms to group the different sub components of RLWE hardware accelerator. All these four algorithms follow different set of rules to find the similarity of the data in the data distribution.
Based on our design requirements, this paper implements four cluster algorithms as stated below: 
\subsection{K-Means Clustering}
K-Means cluster algorithm performs on data to find K divisions to satisfy a certain criteria. Firstly, it computes the distance between the data points and the randomly initialized centroids \cite{kmean}. Thereafter, it creates cluster based on the distance of the data and the centre. This iterative process is repeated until the criteria of the functions converges to the local minimum. Euclidean distance is used to compute distances between the data. Its time complexity is $\mathcal{O}(Kn)$, where $K$ is number of cluster and $n$ is number of data.
As shown in Table \ref{tab:kmean}, the K-Means algorithm creates 5 clusters and each cluster is placed in each FPGA partition. The $V_{ccint}$s of $Partition-1$, $Partition-2$, $Partition-3$, $Partition-4$ and $Partition-5$ are $0.96$, $1.00$, $0.97$, $0.98$ and $0.94$ respectively.
\subsection{DBSCAN}
DBSCAN cluster algorithm works on the assumption that clusters are in high-density regions and outliers tend to be in low-density regions. Unlike the K-Means algorithm, DBSCAN does not require the number of clusters beforehand. This algorithm finds the number of clusters \cite{dbscan} based on two parameters $\bold{epsilon}$ and $\bold{minpoint}$. The $\bold{epsilon:}$ is a radius of the circle around a particular point that is to be considered as in the neighborhood of the other point. The $\bold{minpoint:}$ is the threshold on the least number of points in the circle to be considered as core points. At the first step of DBSCAN choose an arbitrary point $p$. DBSCAN assumes a circle which radious is $\bold{epsilon}$ and centre is at $p$. All the data points comes under this circle is gropued in a cluster. If there are less number of points than the $\bold{minpoint}$, $p$ point is considered as noise. In a newly created cluster, if all points are marked as accessed, then same process used to deal with unvisited points and create a new cluster. This process will continue until all points are marked in cluster or noise. The main advantage of DBSCAN algorithm over the other algorithm is that it can identify the outlier point as a noise. The time complexity of this algorithm is $\mathcal{O}(n)$ for reasonable epsilon. As shown in Table \ref{tab:dbscan}, the DBSCAN algorithm creates 6 clusters. However the offline voltage calibration algorithms stated in Sec. \ref{sec:algo} merge 6 clusters into 3 FPGA partitons, such as (i)\textbf{FPGA Partition-1}: $Cluster-1$, $Cluster-2$ and $Cluster-6$ placed into FPGA Partition-1 where $V_{ccint}$=0.96. (ii)\textbf{FPGA Partition-2}: contain $Cluster-3$ where $V_{ccint}$=0.98. (iii)\textbf{FPGA Partition-3}: $Cluster-4$ and $Cluster-5$ placed into FPGA Partition-3 where $V_{ccint}$=0.97. 

\subsection{Mean-Shift Clustering}
Mean Shift cluster \cite{meanshieft} algorithm is based on the concept of Kernel Density Estimation (KDE). KDE assumes that the data points are sampled from a probability distribution and estimate the distribution by a weight function named as Kernel on each point of the data set. Among many kernels, the Gaussian kernel is the most popular. The mean shift algorithm works in such a way that the points climb uphill to the nearest peak on the KDE surface by iteratively shifting of each point of the data set. It starts with a selected random point as a center of the kernel. It considers a circle which has certain radius in the data set. The kernel is then moved towards a higher density region by shifting the centroid towards the mean of the points within the said circle. The mean shift cluster does not need prior knowledge of the number of clusters. It needs only one parameter: bandwidth to determine the number of clusters. This clustering algorithm is computationally expensive compared to K-means algorithm and its time complexity is $\mathcal{O}(n*log(n))$. As shown in Table \ref{tab:meanshift}, the K-Means algorithm creates 4 clusters and each cluster is placed in each FPGA partition. The $V_{ccint}$s of $Partition-1$, $Partition-2$, $Partition-3$ and $Partition-4$  are $0.96$, $1.00$, $0.97$ and $0.98$ respectively.
\subsection{Hierarchical Clustering}
At the first step, Hierarchical cluster \cite{hierarchical} algorithm considers every point as different clusters. Thereafter it computes distance matrix between two clusters based on a distance measurement method (in this case Euclidean distance). Then it merges two clusters which have the smallest distance. This process will repeat until all clusters grouped into a single cluster. The dendogram creates a binary tree for visualizing the hierarchy of clusters. The number of clusters can be determine from the dendogram. This algorithm is computationally expensive for large datasets, having a time complexity of $\mathcal{O}(n^3)$ where $n$ is the number of data-points. As shown in Table \ref{tab:hierarchical}, the Heirarchical Cluster algorithm creates 6 clusters. However the offline voltage calibration algorithms stated in Sec. \ref{sec:algo} merge 5 clusters into 3 FPGA partitons, such as (i)\textbf{FPGA Partition-1}: $Cluster-1$, $Cluster-5$ placed into FPGA Partition-1 where $V_{ccint}$=0.96.(ii) \textbf{FPGA Partition-2}: $Cluster-2$ and $Cluster-4$ placed into FPGA Partition-2 and fall into same voltage island where $V_{ccint}$=0.98. (iii) \textbf{FPGA Partition-3}: contain $Cluster-3$ where $V_{ccint}$=0.97.  

\begin{figure} [H]
    \centering
    \subfloat[K-Means Clustering]{{\includegraphics[scale=0.5]{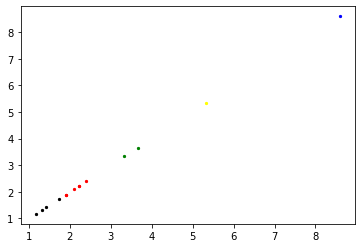} }}
    \qquad
    \subfloat[DBSCAN Clustering]{{\includegraphics[scale=0.5]{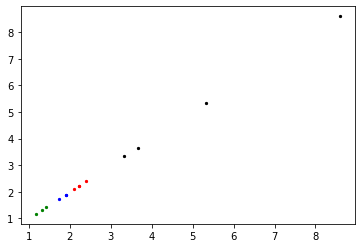} }}
    \qquad \centering
    \subfloat[Mean-Shieft Clustering]{{\includegraphics[scale=0.5]{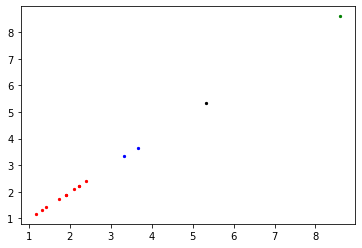} }}
    \label{fig:meanshift}
    \qquad
    \subfloat[Hierarchical Clustering]{{\includegraphics[scale=0.5]{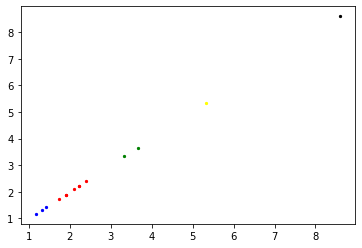} }}
    \caption{Clustering Algorithms}
    \label{fig:algo}
\end{figure}
\section{Voltage Calibration and ErrorControl Unit}
The most challenging part of NTC is (i)Actual voltage allocation for specific FPGA partition and (ii)Detect run time timing errors.
\subsection{Offline Voltage Calibration Algorithm}
\label{sec:algo}
The proposed voltage scaled RLWE has 14 hardware components. These hardware components have different critical paths. The 4 cluster algorithms mentioned in Sec. \ref{sec:cluster} group these 14 components in different voltage island. Each FPGA partition has different biasing voltage $V_{CCint}$. The designer knows if the critical path of a hardware components is more, it needs more amount of $V_{CCint}$ to avoid timing failure. However, designer does not know the actual amount of $V_{CCint}$ required for particular length of critical path. This calibration of $V_{CCint}$ depends on several design constraints, such as number of partition $P$, critical path $C_i$ and critical region ($V_{ccintmax}-V_{ccintmin}$) as shown in Fig. \ref{fig:region}. This Offline Voltage Calibration Algorithm has two parts: (i)Average Critical Path of each Partition shown in Algorithm \ref{alg:algo1} (ii)Voltage Scaling of each Partition shown in \ref{alg:algo2}.
\par
Calculation of average critical path $CP_k$ describes in Algorithm. \ref{alg:algo1}. Total critical path $A_{critical}$ of each partition is computed in Line 4 of Algorithm. \ref{alg:algo1}. Thereafter, Line 6 calculates the average critical path $CP_k$ of $k^{th}$ partition based on $A_{critical}$ and number of components $N(k)$ in $k^{th}$ partition. 
The distributed scaled voltage  $VP_{k}$ for each partition is computed by Algorithm. \ref{alg:algo2}. It calculates voltage $V_{r/c}$ for per unit critical path from $V_{ccintmax}$ and $V_{ccintmin}$ as shown in Line 5. After that, the biasing voltage  $VP_{k}$ of the $k^{th}$ partition is calculated depends on the $V_{r/c}$, average critical path $CP_k$ and $V_{ccintmin}$. The entire RLWE starts running with the calculated biasing voltages $VP_{k}$ for its different FPGA partition. As the $VP_{k}$ falls in the critical region shown in \ref{fig:region}, the timing errors may occurred.
\subsection{Timing Errors}
\begin{figure}[h]
\centering
\begin{minipage}{0.5\textwidth}
\includegraphics[scale=0.34]{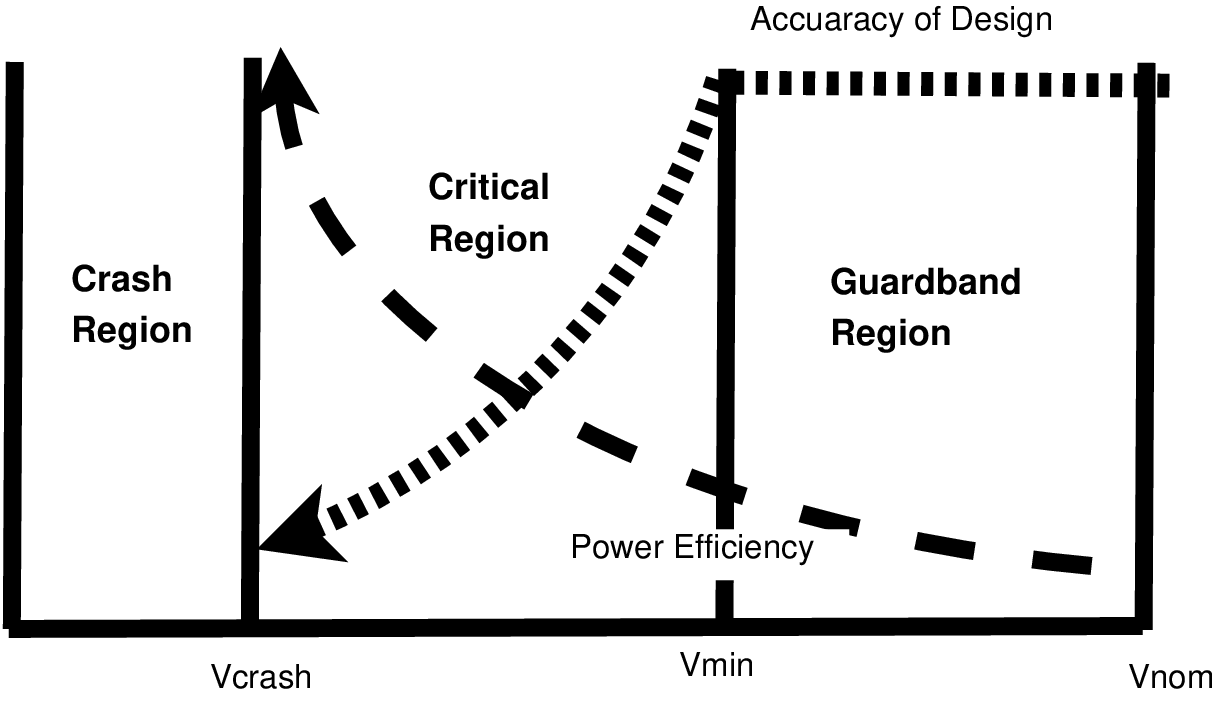}
\vspace{-5pt}
\caption{Voltage behaviour for $V_{CCint}$} \label{fig:region}
\end{minipage}%
\hfil
\begin{minipage}{0.5\textwidth}
\centering
\includegraphics[scale=0.38]{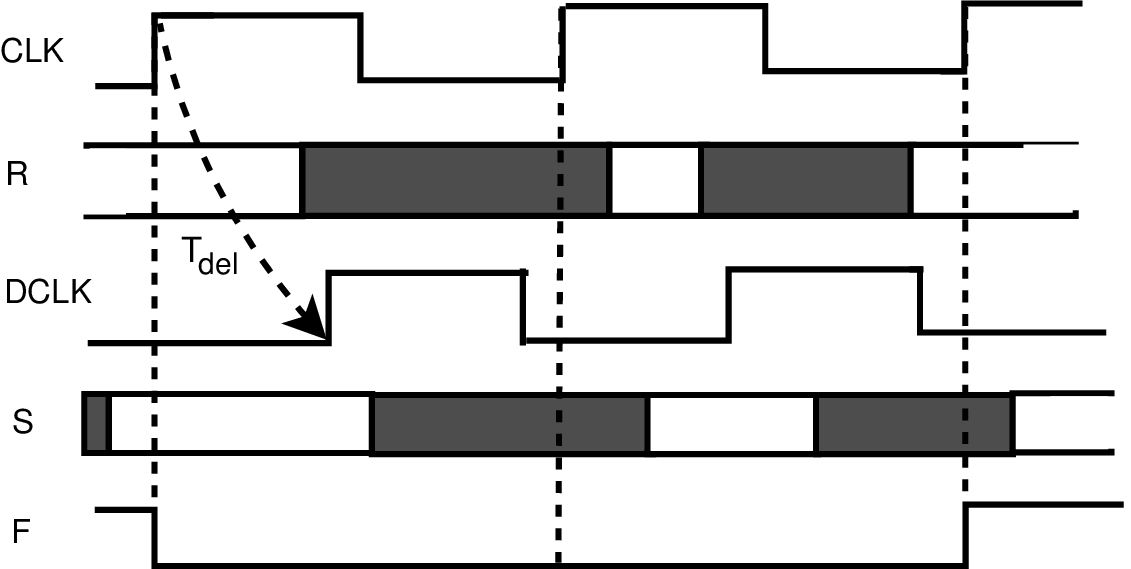}
\vspace{-5pt}
\caption{Razor Timing Diagram of Fault Detection} \label{fig:razor}
\end{minipage}
\end{figure}

Despite of accurate $V_{CCint}$ allocation, NTC on RLWE may causes timing errors. This paper designs a timing error control unit which uses razor flip in the datapath of each RLWE subcomponents. The razor or shadow flipflop in FPGA platform is driven by a delayed clock \cite{razor}. We have assumed that one or more timing pathways leading from any of the source registers terminate at a circuit register $R$. The shadow register $S$ samples the same data as the main register $R$, but it does so on a delayed clock $DCLK$ that is delayed from $CLK$ by $T_{del}$. Any data that enters the circuit after $R$ samples but prior to $S$ samples will result in a disparity between the two registers, which is identified by the error flag $F$. This razor flipflop is put in the datapaths of the subcomponents. Razor increases the resource consumption, but it also has the ability to detect runtime timing errors in RLWE caused by near-threshold biassing voltage. Fig.\ref{fig:razor} displays the timing diagram for the razor. If Razor finds any error in particular partition of the FPGA the biasing voltage of that partition will be increased by one step. Following the method suggested in \cite{boost}, the voltage boosting circuit can be constructed externally.

\begin{table}[H]
\vspace{-10pt}
\caption{K-Means Cluster}
    \label{tab:kmean}
    \vspace{-10pt}
    \centering
    \begin{tabular}{|c|c|c|c|c|}
    \hline
      Cluster  & Hardwares & Critical Paths(ns) & $V_{ccint}(Volt)$&FPGA Partition\\
    \hline
       1      & Dual port ram:1 &                 2.217 &         &\\ 
       & Dual port ram:2       &                  2.217 &         &\\ 
       & Row Column Controller &                   1.9 &    0.96    &1 \\ 
       & Distance              &                  1.9 &         &\\ 
       & reader Wrapper        &                  2.394 &        & \\ 
       & Poly\_add               &                  2.101 &        & \\ 
    \hline
       2     & Datapath        &                  8.603 &   1.00    & 2 \\
    \hline
       3      & Scanner        &                 3.332 &    0.97     &3\\
              & ROM &        3.654 &        &\\   \hline
       4      & Polynomial multiplier  &        5.327 &    0.98     &4\\ \hline
       5      & NTT Controller           &        1.17 &         &\\
              &Convolution Controller   &        1.725 &   0.95    &  5\\
              &Poly\_div                  &        1.323 &        & \\
              &rand                     &        1.412 &         &\\ \hline
    \end{tabular}
    
\end{table}

\begin{table}[H]
\vspace{-10pt}
\caption{DBSCAN Cluster}
    \label{tab:dbscan}
    \vspace{-10pt}
    \centering
    \begin{tabular}{|c|c|c|c|c|}
    \hline
       Clusters  & Hardwares & Critical Paths(ns) & $V_{ccint}(Volt)$& FPGA Partitons\\
    \hline
       1      & Dual port ram:1 &                 2.217 &        &  \\ 
       & Dual port ram:2       &                  2.217 &    0.96 &  1  \\ 
       & reader Wrapper        &                  2.394 &         &\\ 
       & Poly\_add               &                  2.101 &         &\\ \hline
    
       2  & Row Column Controller &                   1.9 &        &\\ 
          & Distance              &                  1.9 &     0.96    & 1\\ 
          &Convolution Controller   &               1.725 &         &\\ \hline
       3     & Datapath        &                  8.603 &   0.98      &2\\  \hline
    
       4      & Scanner        &                 3.332 &    0.97     &3\\
              & ROM &        3.654 &       & \\   \hline
              
       5      & Polynomial multiplier  &        5.327 &    0.97     &3\\ \hline
       
       6      & NTT Controller           &        1.17 &         &\\
              &Poly\_div                 &        1.323 &     0.96  &1  \\
              &rand                     &        1.412 &         &\\ \hline
    \end{tabular}
    
\end{table}

\begin{table}[H]
\vspace{-10pt}
\caption{Mean-Shift Cluster}
    \label{tab:meanshift}
    \vspace{-10pt}
    \centering
    \begin{tabular}{|c|c|c|c|c|}
    \hline
      Cluster & Hardwares & Critical Paths(ns) & $V_{ccint}(Volt)$&FPGA Partition \\
    \hline
       1      & Dual port ram:1 &                 2.217 &         &\\ 
       & Dual port ram:2       &                  2.217 &         &\\ 
       & Row Column Controller &                   1.9 &      &  \\ 
       & Distance              &                  1.9 &         &\\ 
       & reader Wrapper        &                  2.394 &   0.96      & 1\\ 
       & Poly\_add               &                  2.101 &       &  \\ 
       & NTT Controller        &                  1.17 &         &\\
       &Convolution Controller  &                1.725 &          & \\
       &Poly\_div                &               1.323 &        & \\
       &rand                     &              1.412 &         &\\ \hline
    
       2     & Datapath        &                  8.603 &   1.00 &   2  \\
    \hline
       3      & Scanner        &                 3.332 &    0.97  &  3 \\
              & ROM &        3.654 &        &\\   \hline
              
       4      & Polynomial multiplier  &        5.327 &    0.98    & 4\\ \hline
    
    \end{tabular}
    
\end{table}

\begin{table}[H]
\vspace{-10pt}
\caption{Hierarchical}
    \label{tab:hierarchical}
   \vspace{-10pt}
    \centering
    \begin{tabular}{|c|c|c|c|c|}
    \hline
      Cluster  & Hardwares & Critical Paths(ns) & $V_{ccint}(Volt)$& FPGA Partition\\
    \hline
       1      & Dual port ram:1 &                 2.217 &         &\\ 
       & Dual port ram:2       &                  2.217 &        & \\ 
       & Row Column Controller &                   1.9 &        & \\ 
       & Distance              &                  1.9 &       0.96  &1\\ 
       & reader Wrapper        &                  2.394 &        & \\ 
       & Poly\_add               &                  2.101 &       &  \\ 
       &Convolution Controller   &                1.725 &         &\\
    \hline
       2     & Datapath        &                  8.603 &   0.98   &  2\\
    \hline
       3      & Scanner        &                 3.332 &    0.97    & 3\\
              & ROM &        3.654 &        &\\   \hline
              
       4      & Polynomial multiplier  &        5.327 &    0.98     &2\\ \hline
       
       5      & NTT Controller           &        1.17 &        & \\
              &Poly\_div                   &        1.323 &     0.96   &1 \\
              &rand                     &        1.412 &        & \\ \hline
    \end{tabular}
\end{table}

\begin{minipage}{0.46\textwidth}
\begin{algorithm}[H]
\caption{Average Critical Path of each Partition}
\label{alg:algo1}
\begin{algorithmic}[1]
\Require $n_{k}'s,P $   \Comment{$k \gets 0,1,2,....P$}
\State $A_{Critical} \gets 0$
\For{$k \gets 1$ to $P$}                          
       \For{$i \gets 1$ to $N(k)$}                 
            \State {$A_{Critical}$ $\gets$ {$A_{Critical} + C_{i}$}}   
        \EndFor
        \State {$CP_{K} \gets A_{Critical}/N(k) $}
\EndFor
\end{algorithmic}
\end{algorithm}
\end{minipage}
\hfill
\begin{minipage}{0.46\textwidth}
\begin{algorithm}[H]
\caption{Voltage Scaling of each Partition}
\label{alg:algo2}
\begin{algorithmic}[1]
\Require $V_{ccint_{max}},V_{ccint_{min}},CP_{K}'s $  \Comment{$k \gets 0,1,2,....P$}
\State $T_{Critical} \gets 0$
\For{$k \gets 1$ to $P$}                         
      \State {$T_{Critical}  \gets T_{Critical}+CP_{K} $}
\EndFor
\State $V_{r/c} \gets \frac{V_{ccint_{max}}-V_{ccint_{min}}}{T_{Critical}}$
\For{$k \gets 1$ to $P$}                           
      \State {$VP_{k}  \gets V_{ccint_{min}}+CP_{K}*V_{r/c} $}  
\EndFor
\end{algorithmic}
\end{algorithm}
\end{minipage}

\section{Result and Implementation}
\label{sec:rai}
As stated in Sec. \ref{sec:cluster}, 4 cluster algorithms are implemented by using Scikit-learn library of python. The entire framework uses two environments (i) Commercial Vivado tool with Artix-7 FPGA (ii) VTR tool for 22nm, 45nm and 130nm academic FPGAs. This paper proposes two different architectures of RLWE encryption \cite{roy2014compact} scheme for the parameter set (n, q, s) : (256, 7681, 11.32). The first approach implements RLWE hardware accelerator without voltage scaling, whereas the second RLWE architecture is designed with different voltage partitions. Presently, there is no FPGA which supports multiple $V_{ccint}$. Therefore, the power results are carried out separately where one partition is considered as an individual circuit. Both of these two designs should only be studied for critical region as shown in Fig. \ref{fig:region}. However the design and tool constraints have forced us to study both the designs for critical region as well as guard band region. The $V_{ccint}$ of each partitions are calculated by algorithm \ref{alg:algo1} and \ref{alg:algo2}.
\subsection{Vivado and VTR: Voltage Region: $V_{ccint_{min}}=$ 0.95 volt to $V_{ccint_{max}}=$1.05 volt}
The guard band voltage region of Artix-7 is $V_{ccint_{min}}=$ 0.95 volt to $V_{ccint_{max}}=$1.05 volt. Though academic FPGAs on VTR platform supports voltages of critical region, for sake of better comparative study we have implemented 14 components of RLWE with voltage range of $V_{ccint_{min}}=$ 0.95 volt to $V_{ccint_{max}}=$1.05 volt for both $Vivado$ and $VTR$. Table \ref{tab:vivado:vtr} shows $Vivado$ 28nm Artix-7, $VTR$ 22nm, $VTR$ 45nm and $VTR$ 130nm can save 4.34\%, 1.19\%, 1.18\% and 1\% power respectively. 
\subsection{VTR: Voltage Region: $V_{ccint_{min}}=$ 0.5 volt to $V_{ccint_{max}}=$1.3 volt}
Unlike $ VTR$, $Vivado$ does not allow any $V_{ccint}$ below $0.95$ and above $1.05$. As shown in Table \ref{tab:table2}, different number of clusters generated from K-Means, Mean-Shift, DBScan and Hierarchical algorithms consider voltage region from $V_{ccint_{min}}=$ 0.5 volt to $V_{ccint_{max}}=$1.3 volt for 22nm, 45nm and 130nm in VTR flow.

\begin{table}[!htbp]
\caption{Power Consumption of Vivado and VTR : Voltage Region: $V_{ccint_{min}}=$ 0.95 volt to $V_{ccint_{max}}=$1.05} 
\label{tab:vivado:vtr}
\vspace{-10pt}
\centering  
\resizebox{11cm}{!}{%
    \begin{tabular}{|c|c|c|c|c|c|c|c|}
        \hline
Our Design & Cluster     & Partition  & $ V_{ccint_{i}} $ & Vivado& VTR & VTR & VTR \\
 Under   &algorithms    & No. & Volt   &   28nm  & 22nm    & 45nm & 130nm \\  
 $25^{\circ}$ Ambient & & & & & & &  \\
 Temparature & & & & & & &  \\
 \& 100 MHZ Clock & & & & & & &  \\
\hline
Without &     &         &        &        &        &        &\\
Voltage & NA&NA      &   1.00     &    22.45   &   46.44
     &  59.09   & 190.38 \\
scaling& &                &      &         &       &       &\\ \hline
Voltage &        & Partition-1 &   0.96    &       &       &      &\\
scaled&  &         Partition-2 &    1.00    &       &       &      &\\ 
      & K-Means  & Partition-3 &     0.97 &    21.48   &  45.89
     & 58.402  & 188.49 \\
      &         &   Partition-4 &   0.98   &       &       &        & \\
     &         &   Partition-5 &    0.95  &       &       &        & \\
 \hline
   \multicolumn{4}{|c|}{\textbf{ \% of Reduction}} & 4.34 &1.19  & 1.18 & 1.005 \\
 \hline
 
Without &     &         &        &        &        &        &\\
Voltage &NA&NA      &   1.00     &   22.45
     &    46.44    & 59.09        & 190.38 \\
scaling& &                &      &         &       &       &\\ \hline
Voltage &        & Partition-1 &    0.96   &       &       &      &\\
scaled&  &         Partition-2 &     0.97   &  21.62     &   45.94    & 58.46     & 187.2 \\ 
      & Mean-Shift & Partition-3 &   1   &       &       &        &\\
      &         &   Partition-4 &  0.98    &       &       &        & \\
 \hline
  \multicolumn{4}{|c|}{\textbf{ \% of Reduction}} & 3.73 &1.1  & 1.09 & 1.69 \\
 \hline
 Without &     &         &        &        &        &        &\\
Voltage & NA&NA      &  1.00      &    22.45    &    46.44    &59.09        & 190.38 \\
scaling& &                &      &         &       &       &\\ \hline
Voltage &        & Partition-1 &   0.96    &       &       &      &\\
scaled&  &         Partition-2 &     0.96   &       &       &      &\\ 
      & DBSCAN & Partition-3 &   0.97   &    21.16   &   45.87    &58.38        & 188.5 \\
      &         &   Partition-4 &    0.98  &       &       &        & \\
 \hline
 \multicolumn{4}{|c|}{\textbf{ \% of Reduction}} & 5.78 &1.24  & 1.21 & 0.99 \\
 \hline
Without &     &         &        &        &        &        &\\
Voltage & NA&NA      &    1.00    &    22.45    &    46.44    & 59.09       & 190.38 \\
scaling& &                &      &         &       &       &\\ \hline
Voltage &        & Partition-1 &    0.96   &       &       &      &\\
scaled&  &         Partition-2 &    0.96    &       &       &      &\\ 
  & Hierarchical& Partition-3 &   0.97   &   21.62    & 45.94      &  58.46    & 188.6 \\
    &         &   Partition-4 &   0.98  &       &       &        & \\
     &         &   Partition-5 &   1   &       &       &        & \\
 \hline
 \multicolumn{4}{|c|}{\textbf{ \% of Reduction}} & 3.73 &1.1  & 1.09 & 0.93 \\
 \hline
    \end{tabular}
}
\vspace{-10pt}
\end{table}

\begin{table}[H]
\caption{Power Consumption for VTR, Voltage Region: $V_{ccint_{min}}=$ 0.5 volt to $V_{ccint_{max}}=$1.3} 
\label{tab:table2}
\vspace{-10pt}
\centering  
\resizebox{11cm}{!}{%
    \begin{tabular}{|c|c|c|c|c|c|c|}
        \hline
Our Design & Cluster     & Partition  & $ V_{ccint_{i}} $ &  VTR & VTR & VTR \\
 Under   &algorithms    & No. & Volt   &    22nm    & 45nm & 130nm \\  
 $25^{\circ}$ Ambient & & & & & &   \\
 Temperature & & & & & &   \\
 \& 100 MHZ Clock & & & & & &   \\
\hline
Without &     &         &        &        &        &        \\
Voltage & NA&    NA  &   1.00     &       46.44
     &  59.09   & 190.38   \\
scaling& &                &      &         &             &\\ \hline
Voltage &        & Partition-1 &   0.96    &              &      &\\
scaled&  &         Partition-2 &    1.00    &              &      &\\ 
      & K-Means  & Partition-3 &     0.97 &      41.41 & 52.69  & 179.55 \\
      &         &   Partition-4 &   0.98   &              &        & \\
     &         &   Partition-5 &    0.95  &             &        & \\
 \hline
   \multicolumn{4}{|c|}{\textbf{ \% of Reduction}} & 10.83  & 10.82 & 5.68 \\
 \hline
 
Without &     &         &        &        &               &\\
Voltage & NA&NA      &   1.00     
     &    46.44    & 59.09        & 190.38 \\
scaling& &                &               &       &       &\\ \hline
Voltage &        & Partition-1 &    0.96   &              &      &\\
scaled&  &         Partition-2 &     0.97     &   41.71    & 53.28     & 181.2 \\ 
      & Mean-Shift & Partition-3 &   1   &              &        &\\
      &         &   Partition-4 &  0.98    &              &        & \\
 \hline
  \multicolumn{4}{|c|}{\textbf{ \% of Reduction}} & 10.19  & 9.83 & 4.82 \\
 \hline
 Without &     &         &                &        &        &\\
Voltage & NA&    NA  &  1.00         &    46.44    &59.09        & 190.38 \\
scaling& &                &               &       &       &\\ \hline
Voltage &        & Partition-1 &   0.96    &             &      &\\
scaled&  &         Partition-2 &     0.96   &              &      &\\ 
      & DBSCAN & Partition-3 &   0.97   &       41.27    &52.77        & 178.2 \\
      &         &   Partition-4 &    0.98  &       &       &         \\
 \hline
 \multicolumn{4}{|c|}{\textbf{ \% of Reduction}}  &11.15  & 10.71 & 6.38 \\
 \hline
Without &     &         &        &        &                &\\
Voltage & NA&NA      &    1.00        &    46.44    & 59.09       & 190.38 \\
scaling& &                &      &         &              &\\ \hline
Voltage &        & Partition-1 &    0.96   &              &      &\\
scaled&  &         Partition-2 &    0.96    &              &      &\\ 
  & Hierarchical& Partition-3 &   0.97       & 41.48      &  53    & 181.1 \\
    &         &   Partition-4 &   0.98  &              &        & \\
     &         &   Partition-5 &   1   &              &        & \\
 \hline
 \multicolumn{4}{|c|}{\textbf{ \% of Reduction}} &10.69  & 10.32 & 4.89 \\
 \hline
    \end{tabular}
     }
\end{table}

\begin{table}[H]
\vspace{-10pt}
\caption{Comparison of our design with other RLWE encryption schemes} 
\vspace{-10pt}
\label{tab:compare}
\centering  
  \resizebox{11cm}{!}{
   \begin{tabular}{|c|c|c|c|c|c|c|c|c|c|}
      \hline
Designs & Parameters & Device & Slice    & Slice & DSP & BRAM & Critical & Throughput & Year\\
        &(n,q,s)     &        & Register/FF & LUT   &     &      & Path (NS)&(bps)     &        \\
   \hline

Our     &  (256,7681,11.32)  & Artix-7       & 665/309    & 846      &  2    & 1   &2.539 &     $\sim 7.2\times 10^{6}$ & 2022   \\
Design  &            &        &          &       &     &      &           &             & \\
\hline
RLWE & (256,7681,11.32)& V6LX75T &1506/3624& 4549& 1 & 12   &     -      &  $\sim 9.77\times 10^{6}$            &\\
      &            & @262 MHz   &          &       &     &      &           &             & \\

\cite{pop} &  (512,12289,12.18)&V6LX75T &1887/4760& 5595& 1& 14   &     -      &    $\sim 9.33\times 10^{6}$   &2013\\
      &            & @251 MHz   &          &       &     &      &           &             & \\
\hline
RLWE& (256,7681,11.32)& V6LX75T &-/860& 1349& 1 & 2   &     -      &  $\sim 12.74\times 10^{6}$           & \\
      &            & @313 MHz   &          &       &     &      &           &             & \\

\cite{roy2014compact}& (512,12289,12.18)& V6LX75T &-/953& 1536& 1 & 3   &     -      &  $\sim 10.69\times 10^{6}$           & 2014\\
      &            & @278 MHz   &          &       &     &      &           &            &  \\
\hline
\cite{gottert2012design}& (256,7681,11.32) &V6LX240T & 143396/- & 298016 & - & -      & -&  $\sim 31.8\times 10^{6}$ &2012\\
      &            &      &          &       &     &      &           &             & \\

\hline
\cite{hal}-V1&    (256,7681,11.32)   &   XC7A200    &1624/-       & 4365   & 1  &  5   &     -      &     $\sim 10.58\times 10^{6}$ & 2019\\
      &               & @208 MHz   &          &       &     &      &           &              &\\
\hline
\cite{hal}-V2&   (256,7681,11.32)   &  XC7A200      &2122/-       & 5616   & 6  &  8    &     -     &  $\sim 16.95\times 10^{6}$  &2019\\
      &             & @250 MHz   &          &       &     &      &           &             & \\
\hline
\cite{Howe}& (256,4096,3.39) & S6LX45& 1866/4804 & 6152 & 1 & 73      & -&  $\sim 0.32\times 10^{6}$ &2016\\
      &             & @125 MHz   &          &       &     &      &           &              &\\
\hline
    \end{tabular}
    }
\end{table}

   As an example, K-Means algorithm generates five partitions. Algorithm \ref{alg:algo1} calculates average critical path $CP_{k}$ of the $k^{th}$ partition where $A_{Critical}$ is the total critical path the $k^{th}$ partition. Then algorithm \ref{alg:algo2} first computes the voltage per unit critical path $V_{r/c}$, after that required biasing voltage $VP_{k}$ for $k^{th}$ partition is calculated in line 7 of algorithm \ref{alg:algo2}. The five $V_{ccint}$s calculated by algorithm \ref{alg:algo2} using K-Means are : $V_{ccint_{1}}=0.96$ $V_{ccint_{2}}=0.995 \approx 1 $, $V_{ccint_{3}}=0.965 \approx 0.97$ $V_{ccint_{4}}=0.975 \approx 0.98 $, $V_{ccint_{5}}=0.955 \approx 0.95$. In Table \ref{tab:vivado:vtr}, our voltage scaling method in same voltage ranges, reduces the dynamic power consumption $3.73 \% $ to $5.78 \%$ in $Vivado$ environment and reduces $0.93 \%$ to $1.24 \%$ in VTR environment.Table \ref{tab:table2} shows adoption of voltage scaling technology reduces dynamic power consumption $10.19 \% $ to $11.15 \%$, $9.83 \% $ to $10.82 \%$ and $4.82 \% $ to $6.38 \%$ for 22nm, 45nm and 130nm VTR flow respectively. Table \ref{tab:compare} shows comparison of resource and throughput of our RLWE hardware accelerator with existing literature. Though our paper does not claim any architectural contribution, Table \ref{tab:compare} reports that resource usage and throughput are significantly better compared to existing RLWE.

\section{Conclusion}
\label{sec:con}
This paper implements a low power RLWE hardware accelerator in Artix-7 28nm commercial FPGA , 22nm, 45nm and 130 nm academic FPGA using $Vivado$ and $VTR$ tool. The proposed methodology creates clusters of subcomponents of RLWE based on its critical path. The FPGA floor is partitioned and different clusters are placed in different FPGA partitions. The clusters which have higher average critical path is connected with higher biasing voltage and the clusters which have lower average critical path is connected with lower biasing voltage to avoid the timing failure. This paper proposes an algorithm to calculate the actual biasing voltage required for certain amount of average critical path of a partition based on the available FPGA technology ($V_{ccint_{max}},V_{ccint_{min}}$) and number of partitions. The Razor flipflop used in each subcomponents of RLWE can detect any timing error causes for NTC computation. This voltage scaled, partitioned RLWE can save $\sim$6\% and $\sim$11\% power in Vivado and VTR platform respectively. The resource usage and throughput of the implemented RLWE hardware accelerator is comparatively better than existing literature. 
Full version of this manuscript is available in IEEE Access \cite{rourabieee}.

\bibliographystyle{unsrt}
\bibliography{Journal}

\begin{thebibliography}{10}

\bibitem{grover}
Markus Grassl, Brandon Langenberg, Martin Roetteler, and Rainer Steinwandt.
\newblock Applying grover’s algorithm to aes: quantum resource estimates.
\newblock In {\em Post-Quantum Cryptography}, pages 29--43. Springer, 2016.

\bibitem{regev11}
Oded Regev.
\newblock On lattices, learning with errors, random linear codes, and
  cryptography.
\newblock In {\em Proceedings of the thirty-seventh annual ACM symposium on
  Theory of computing}, pages 84--93, 2005.

\bibitem{micciancio2009}
Daniele Micciancio and Oded Regev.
\newblock Lattice-based cryptography.
\newblock In {\em Post-quantum cryptography}, pages 147--191. Springer, 2009.

\bibitem{pop}
Thomas P{\"o}ppelmann and Tim G{\"u}neysu.
\newblock Towards practical lattice-based public-key encryption on
  reconfigurable hardware.
\newblock In {\em International Conference on Selected Areas in Cryptography},
  pages 68--85. Springer, 2013.

\bibitem{gottert2012design}
Norman G{\"o}ttert, Thomas Feller, Michael Schneider, Johannes Buchmann, and
  Sorin Huss.
\newblock On the design of hardware building blocks for modern lattice-based
  encryption schemes.
\newblock In {\em International Workshop on Cryptographic Hardware and Embedded
  Systems}, pages 512--529. Springer, 2012.

\bibitem{poppelmann2014}
Thomas P{\"o}ppelmann and Tim G{\"u}neysu.
\newblock Area optimization of lightweight lattice-based encryption on
  reconfigurable hardware.
\newblock In {\em 2014 IEEE international symposium on circuits and systems
  (ISCAS)}, pages 2796--2799. IEEE, 2014.

\bibitem{sinhaKnuth}
Sujoy Sinha~Roy, Frederik Vercauteren, and Ingrid Verbauwhede.
\newblock High precision discrete gaussian sampling on fpgas.
\newblock In {\em International Conference on Selected Areas in Cryptography},
  pages 383--401. Springer, 2013.

\bibitem{aysu2013low}
Aydin Aysu, Cameron Patterson, and Patrick Schaumont.
\newblock Low-cost and area-efficient fpga implementations of lattice-based
  cryptography.
\newblock In {\em 2013 IEEE international symposium on hardware-oriented
  security and trust (HOST)}, pages 81--86. IEEE, 2013.

\bibitem{hybrid}
Paul~S Zuchowski, Christopher~B Reynolds, Richard~J Grupp, Shelly~G Davis,
  Brendan Cremen, and Bill Troxel.
\newblock A hybrid asic and fpga architecture.
\newblock In {\em IEEE/ACM International Conference on Computer Aided Design,
  2002. ICCAD 2002.}, pages 187--194. IEEE, 2002.

\bibitem{Kusse}
E.~Kusse and J.~Rabaey.
\newblock Low-energy embedded fpga structures.
\newblock In {\em Proceedings. 1998 International Symposium on Low Power
  Electronics and Design (IEEE Cat. No.98TH8379)}, pages 155--160, 1998.

\bibitem{Howe}
J.~Howe, C.~Moore, M.~O'Neill, F.~Regazzoni, T.~Güneysu, and K.~Beeden.
\newblock Lattice-based encryption over standard lattices in hardware.
\newblock In {\em 2016 53nd ACM/EDAC/IEEE Design Automation Conference (DAC)},
  pages 1--6, 2016.

\bibitem{roy2014compact}
Sujoy~Sinha Roy, Frederik Vercauteren, Nele Mentens, Donald~Donglong Chen, and
  Ingrid Verbauwhede.
\newblock Compact ring-lwe cryptoprocessor.
\newblock In {\em International workshop on cryptographic hardware and embedded
  systems}, pages 371--391. Springer, 2014.

\bibitem{poppelmann2014enhanced}
Thomas P{\"o}ppelmann, L{\'e}o Ducas, and Tim G{\"u}neysu.
\newblock Enhanced lattice-based signatures on reconfigurable hardware.
\newblock In {\em International Workshop on Cryptographic Hardware and Embedded
  Systems}, pages 353--370. Springer, 2014.

\bibitem{poppelmann2012}
Thomas P{\"o}ppelmann and Tim G{\"u}neysu.
\newblock Towards efficient arithmetic for lattice-based cryptography on
  reconfigurable hardware.
\newblock In {\em International conference on cryptology and information
  security in Latin America}, pages 139--158. Springer, 2012.

\bibitem{lindner}
Richard Lindner and Chris Peikert.
\newblock Better key sizes (and attacks) for lwe-based encryption.
\newblock In {\em Cryptographers’ Track at the RSA Conference}, pages
  319--339. Springer, 2011.

\bibitem{Damien}
Damien Stehl{\'e} and Ron Steinfeld.
\newblock Making ntru as secure as worst-case problems over ideal lattices.
\newblock In {\em Annual international conference on the theory and
  applications of cryptographic techniques}, pages 27--47. Springer, 2011.

\bibitem{lyubashevsky1}
Vadim Lyubashevsky, Chris Peikert, and Oded Regev.
\newblock On ideal lattices and learning with errors over rings.
\newblock In {\em Annual international conference on the theory and
  applications of cryptographic techniques}, pages 1--23. Springer, 2010.

\bibitem{trivium}
Turgay {Kaya}.
\newblock {Memristor and Trivium-based true random number generator}.
\newblock {\em Physica A Statistical Mechanics and its Applications},
  542:124071, March 2020.

\bibitem{knuth}
Donald~E Knuth and YAO AC.
\newblock The complexity of nonuniform random number generation.
\newblock 1976.

\bibitem{vtr}
Kevin~E. Murray, Oleg Petelin, Sheng Zhong, Jia~Min Wang, Mohamed Eldafrawy,
  Jean-Philippe Legault, Eugene Sha, Aaron~G. Graham, Jean Wu, Matthew J.~P.
  Walker, Hanqing Zeng, Panagiotis Patros, Jason Luu, Kenneth~B. Kent, and
  Vaughn Betz.
\newblock Vtr 8: High-performance cad and customizable fpga architecture
  modelling.
\newblock {\em ACM Trans. Reconfigurable Technol. Syst.}, 13(2), May 2020.

\bibitem{kmean}
David Arthur and Sergei Vassilvitskii.
\newblock k-means++: The advantages of careful seeding.
\newblock Technical report, Stanford, 2006.

\bibitem{dbscan}
Martin Ester, Hans-Peter Kriegel, J{\"o}rg Sander, Xiaowei Xu, et~al.
\newblock A density-based algorithm for discovering clusters in large spatial
  databases with noise.
\newblock In {\em kdd}, volume~96, pages 226--231, 1996.

\bibitem{meanshieft}
Dorin Comaniciu and Peter Meer.
\newblock Mean shift: A robust approach toward feature space analysis.
\newblock {\em IEEE Transactions on pattern analysis and machine intelligence},
  24(5):603--619, 2002.

\bibitem{hierarchical}
D~Manning Christopher, Raghavan Prabhakar, and Schutze Hinrich.
\newblock Introduction to information retrieval, 2008.

\bibitem{razor}
D.~{Ernst}, {Nam Sung Kim}, S.~{Das}, S.~{Pant}, R.~{Rao}, {Toan Pham},
  C.~{Ziesler}, D.~{Blaauw}, T.~{Austin}, K.~{Flautner}, and T.~{Mudge}.
\newblock Razor: a low-power pipeline based on circuit-level timing
  speculation.
\newblock In {\em Proceedings. 36th Annual IEEE/ACM International Symposium on
  Microarchitecture, 2003. MICRO-36.}, pages 7--18, 2003.

\bibitem{boost}
T.~N. {Miller}, X.~{Pan}, R.~{Thomas}, N.~{Sedaghati}, and R.~{Teodorescu}.
\newblock Booster: Reactive core acceleration for mitigating the effects of
  process variation and application imbalance in low-voltage chips.
\newblock In {\em IEEE International Symposium on High-Performance Comp
  Architecture}, pages 1--12, 2012.

\bibitem{hal}
Timo Zijlstra, Karim Bigou, and Arnaud Tisserand.
\newblock Fpga implementation and comparison of protections against scas for
  rlwe.
\newblock In {\em International Conference on Cryptology in India}, pages
  535--555. Springer, 2019.

\bibitem{rourabieee}
Rourab~Paul Paresh~Baidya, Swagata~Mandal.
\newblock Near threshold computation of partitioned ring learning with error
  ({RLWE}) hardware accelerator on reconfigurable architecture, doi:
  10.1109/access.2024.3401235.
\newblock In {\em IEEE Access}, 2024.

\end{thebibliography}

\end{document}